\documentclass[conference]{IEEEtran}
\IEEEoverridecommandlockouts
\usepackage{cite}
\usepackage{amsmath,amssymb,amsfonts}
\usepackage{algorithmic}
\usepackage{graphicx}

\usepackage{stackengine}
\usepackage{comment}
\usepackage{ulem}
\usepackage{tabularx}

\usepackage{textcomp}
\usepackage{xcolor}
\usepackage[ruled]{algorithm2e}

\def\BibTeX{{\rm B\kern-.05em{\sc i\kern-.025em b}\kern-.08em
    T\kern-.1667em\lower.7ex\hbox{E}\kern-.125emX}}
\newcommand*{\Perm}[2]{{}^{#1}\!P_{#2}}%
\usepackage{fancyhdr}
\begin{document}

\title{QoS-Aware Load Balancing in Wireless Networks using Clipped Double Q-Learning \\

}

\author{\IEEEauthorblockN{ Pedro Enrique Iturria-Rivera, \IEEEmembership{Student Member,~IEEE}, Melike Erol-Kantarci, \IEEEmembership{Senior Member,~IEEE}}
\IEEEauthorblockA{\textit{School of Electrical Engineering and Computer Science}
\textit{University of Ottawa}\\
Ottawa, Canada \\
\{pitur008, melike.erolkantarci\}@uottawa.ca}}

\maketitle

\begin{abstract}
In recent years, long-term evolution (LTE) and 5G NR (5$^{th}$ Generation New Radio) technologies have showed great potential to utilize Machine Learning (ML) algorithms in optimizing their operations, both thanks to the availability of fine-grained data from the field, as well as the need arising from growing complexity of networks. The aforementioned complexity sparked mobile operators' attention as a way to reduce the capital expenditures (CAPEX) and the operational (OPEX) expenditures of their networks through network management automation (NMA). NMA falls under the umbrella of Self-Organizing Networks (SON) in which 3GPP has identified some challenges and opportunities in load balancing mechanisms for the Radio Access Networks (RANs). In the context of machine learning and load balancing, several studies have focused on maximizing the overall network throughput or the resource block utilization (RBU). In this paper, we propose a novel Clipped Double Q-Learning (CDQL)-based load balancing approach considering resource block utilization, latency and the Channel Quality Indicator (CQI). We compare our proposal with a traditional handover algorithm and a resource block utilization based handover mechanism. Simulation results reveal that our scheme is able to improve throughput, latency, jitter and packet loss ratio in comparison to the baseline algorithms. 

\end{abstract}

\begin{IEEEkeywords}
load balancing, double clipped q-learning, wireless networks.
\end{IEEEkeywords}

\section{Introduction}
The exponential increase in mobile network usage alongside the growing data consumption demand by several use cases such as AR/VR (Augmented Reality/Virtual Reality) and video streaming, have led wireless networks to evolve in order to satisfy the pressing requirements. More specifically, video-dominated applications related to video streaming, video conferencing and high quality buffered media have faced a great spike due the recent impact of the COVID-19 pandemic. Some studies showed that  after the lockdown, an increase of 215-285\% in VoIP and videoconferencing traffic and a 20-40\% increment in streaming and web video consumption have been observed \cite{Lutu2020}. Besides, the bandwidth requirement of the aforementioned use cases, they also have tight delay requirements. The trend for multimedia usage is expected to continue after the pandemic and mobile network operators will shoulder a majority of the load. This calls for optimized use of resources at the RAN, the transport network and the core.     

LTE (Long Term Evolution) and its successor 5G (5th Generation) NR (New Radio) are able to support self-optimization functionalities which 3GPP has identified in the context of SON. These also include load balancing. Load balancing for the RAN, involves handing over UEs (User Equipments) to less occupied base stations. Naturally, a base station's load would be in relation to the number of UEs that are associated to it and the traffic demand of these UEs. 

As part of the handover mechanism, each UE in the network will send periodical measurement reports to its respective serving cell. In practical terms, the serving cell will send via a Radio Resource Control (RRC) message the indication of what type of measurements each connected UE must gather and consequently report of the vicinity cells and of itself.  Then, BSs will look for handover opportunities by verifying some possible events to  trigger or not the handover procedure. Some of the events described by 3GPP are: \cite{3GPP2018}: 

\begin{itemize}
\item A2: Serving cell Reference Signal Received Quality (RSRQ) becomes worse than threshold
\item A3: Neighbour Reference Signal Received Power (RSRP) becomes better than serving cell 
\item A4: Neighbour cell RSRQ becomes better than threshold \end{itemize}

In the context of handover, there are two well-known mechanisms. These are A2-A4 handover where two conditions must be satisfied corresponding to the events A2 and A4: $i)$ to trigger handover the serving cell of the UE must fall below certain RSRQ serving cell threshold and $ii)$ the handover is only performed if the difference between the best neighbor and the serving cell RSRQ is greater than certain predefined neighbor cell offset. The A3 handover algorithm or ``strongest cell handover algorithm" is a simpler approach where handover is triggered for the UE to the best cell in the measurement report. When the best cell in terms of RSRP is selected, this value must be greater than the current serving cell by an hysteresis value and must be maintained for a time (TTT, time to trigger) in order to avoid ping-pong effect. Handover using the previous mechanism is illustrated in Fig. \ref{a3}.

\begin{figure}
  \includegraphics[width=\columnwidth]{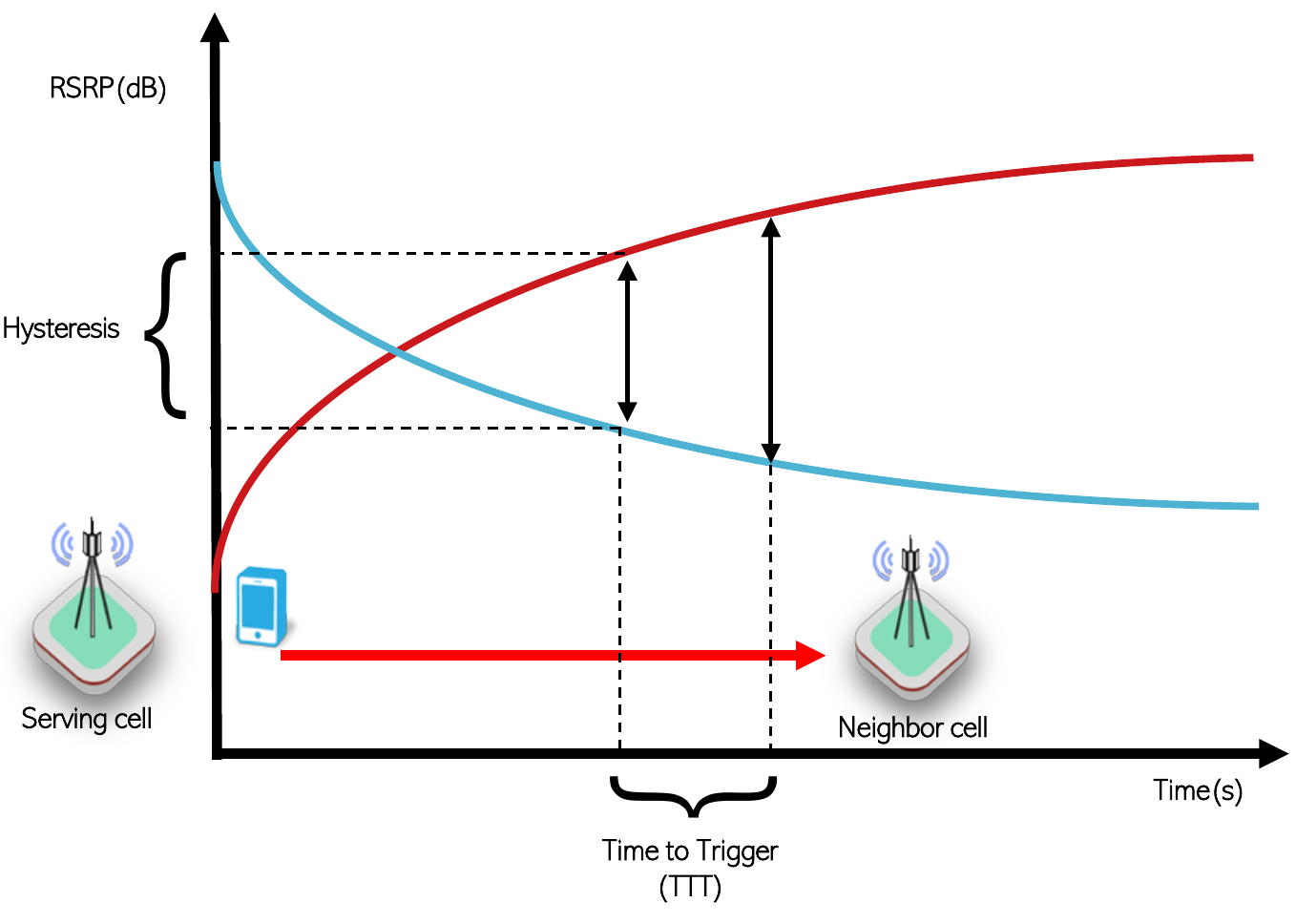}
  \caption{Illustration of A3 handover algorithm.}
  \label{a3}
\end{figure}

In this paper, we address the load balancing problem in wireless networks where we seek to balance resource block utilization and QoS metrics such as latency and throughput. To do so, we use clipped double Q-learning which tackles the overestimation issues presented by previous deterministic action space RL algorithms. We consider a centralized approach where our agent can choose the best cell individual offset (CIO) per BSs that maximizes the key performance indicators of the network. Such a centralized learning can be utilized in Cloud RAN (C-RAN). We compare our results with the classical A3 Handover algorithm and a simplistic resource block utilization-based baseline handover algorithm. Our results show an improvement in the overall performance of the network in terms of throughput, latency, jitter, and packet loss ratio when the proposed algorithm is used. 

This paper is organized as follows. Section II presents literature related to load balancing in wireless networks. In section III, the description of the system model and the motivation of this paper is introduced. Section IV provides a description of the proposed scheme. Section V depicts the performance evaluation and comparison with the baseline algorithms. Finally, section VI concludes the paper.

\section{Related work}
There are several works in the literature related to load balancing in RANs due its significant impact on saving resources and maximizing the network performance. In most of the cases, the latter problem is tackled by modifying the existing handover (HO) strategies either by modifying some HO parameters or by constantly tracking the network Key Performance Indicators (KPIs). \cite{Ahmad2017,Tayyab2019}. 

Recently, machine learning has been used for load balancing as well. In \cite{Yajnanarayana2020}, the authors proposed a 5G handover algorithm based on Q-learning for optimizing the handover mechanism by the selection of data link beams and access beams in 5G cellular networks. In \cite{Huang2020} the authors presented a supervised learning solution based on deep learning by considering the variation of the SINR (Signal to Noise Ratio) to calculate the probability of Radio Link Failure (RLF) and based on this metric, they performed  handover to the cell that is less prone to experience RLF. In \cite{Xu2019} the authors presented an RL-based mobility load balancing (MLB) algorithm addressed to deal with ultra-dense networks. The proposal consisted of a two-layer architecture with the first layer in charge of building small clusters and the second layer where in each intra-cluster, an MLB algorithm is executed to obtain the optimal HO parameters by minimizing the resource block utilization. In \cite{Attiah2020} the authors presented an RL-based load balancing algorithm for LTE where maximizing the instantaneous throughput of the overall network is the main objective. 

In summary, previous works do not consider end-to-end delay and the channel quality in the optimization process of HO parameters. Different than other papers, we consider additional QoS metrics alongside the resource block utilization seeking to balance both while maximizing the network performance.      
\section{System model}
\label{AA}
We consider a network consisting of a set $\Gamma$  of size $M_T$ of base stations (BS). The network serves a set of $\Psi$ of size $N_T$ stationary mobile users deployed randomly around the BSs.  

According the Downlink (DL) bandwidth configuration chosen, $B$ MHz, let us define $N_{RB}^{DL}$ as the number of resource blocks available to be assigned by the Media Access Control (MAC) scheduler. The Channel and QoS Aware (CQA) scheduler is chosen as MAC scheduler. This scheduler assigns resource blocks by prioritizing users with greater Head of Line (HOL) delay  and maximizing  MAC layer throughput\cite{Bojovic2014}. In this paper, we consider one Resource Block Group (RBG) ($RBG = N_{RB}^{DL} / K$) as the smallest resource unit where $K\in[1,2,3,4]$ according the DL bandwidth used. In addition, a centralized approach is assumed where a central agent is able to monitor the BSs and UEs Key Performance Indicators (KPIs). This can be conveniently applicable to C-RANs, as mentioned before. The agent is capable of altering the CIO of each $i$ cell based on the proposed machine learning algorithm. The CIO of each cell denoted as $\o_{i} \in [\o_{min}, \o_{max}]$ dB and is modified in order to trigger the handover algorithm among cells. For A3 handover algorithm, if a user is served by some cell $i$, it will start a handover to cell $j$ request through the $X2$ interface if the following condition holds: 

\begin{equation}
    RSPR_j + \o_{j\rightarrow i} > Hys + RSPR_i + \o_{i\rightarrow j}
    \label{a3}
\end{equation}
where  $RSPR_j$ and $RSPR_i$ are the measured RSRP values in dB of the serving cell and the neighbor cell. $Hys$ is the hysteresis value used to avoid ping-pong scenarios.
$\o_{j\rightarrow i}$ and $\o_{i\rightarrow j}$ are the cell individual offsets. The values are independent for each cell.

\section{Clipped Double Q-Learning Based Load Balancing}
In the proposed approach, an agent observes the environment parameters, such as CQI, packet delay and resource block utilization per BS. The agent's actions consist of modifying the BS's individual CIO values in order to maximize the agent's reward. In the following subsection, we present an overview of the Clipped Double Q-Learning (CDQL) and then formally define our solution.

\subsection{Clipped Double Q-Learning }\label{AA}

In this work, we use a state-of-the-art deterministic action space RL algorithm named Clipped Double Q-Learning. This algorithm is presented in \cite{Fujimoto2018}
as part of TD3 (Twin Delayed Deep Deterministic Policy
Gradient Algorithm), which builds on the Deep Deterministic Policy Gradient algorithm (DDPG) \cite{Lillicrap2016}. It may be seen as a continuation of the work of \cite{VanHasselt2016} with its Double Deep Q-Learning algorithm (DDQN). The aforementioned algorithms suffered from Q-value overestimation due the usage of the $argmax$ operator for selecting the maximum Q-value. CDQL tackles overestimation issues by following the strategy of having two neural networks that learn at the same time meanwhile the reward is calculated based on the minimum Q-value of such networks hence, reducing overestimation. The pseudo-code for CDQL is presented in Algorithm \ref{cddqn}. 
\begin{figure}
  \includegraphics[scale=0.9]{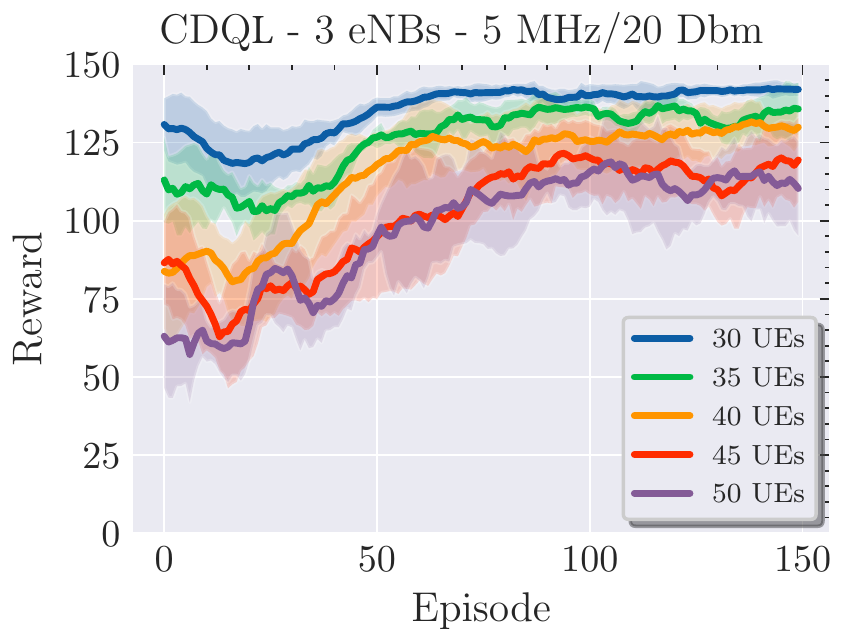}
  \caption{Learning performance for varying number of UEs and for the CDQL algorithm. }
  \label{cddqn_learning}
\end{figure}

\subsection{Action space selection}\label{AA}
The actions of our agent are defined as a vector with the CIO values assigned to each BS. For CDQL the set of actions will be deterministicly predefined by the possible permutations of the set of predefined values. 

For CDQL algorithm, the size of the action space will be $\Perm{l}{k}=\frac{l!}{(l-k)!}$ where $l$ is the size of the set of the possible CIO values that can take each BS  and $k$ will be equivalent to $M_T$ or the amount of BSs in the network. Thus,
\begin{equation}
    A(t) =\begin{bmatrix} \o_{1}(t) &\o_{2}(t)& ...& \o_{M_T}(t) \end{bmatrix}
\end{equation}

The individual CIO value defined as $\o_{i}$ will be lower and upper bounded by two predefined values $\o_{min}$ and $\o_{max}$ as described as follows: $\o_{i}(t) \in [\o_{min}, \o_{max}]$.

\normalem 
\begin{algorithm}
\SetAlgoLined
    
 Initialize $Q_{\theta_1}$,$Q_{\theta_2}$ and target networks with random weights, replay buffer $J$, $\epsilon$, $\epsilon_{decay}$ and $\epsilon_{min}$ \;
 
   \ForEach{environment step}
    {
        Observe state $s_t$\;
        Select $a_t \sim \pi(a_t, s_t)$ \textbf{if} $\epsilon \geq x \sim \emph{U}(0,1)$ \textbf{otherwise} select random action $a_t$ \;
        Execute $a_t$ and observe next state $s_{t+1}$ and reward $r_t= R(s_t, a_t)$\;
        Store $(s_t,a_t, r_t, s_{t+1})$ in replay buffer $J$\;
        \ForEach{\text{update} }{
        sample $e_t = (s_t,a_t, r_t, s_{t+1}) \sim J $\;
        Compute $Y_{t}^{CDQL}$:\ 
            
            $Y_{t}^{CDQL} = r_t + \gamma min_{i= 1,2}Q_{\theta_i}(s_{t+1}, argmax_{a'}$ $Q_{\theta_i}(s_{s+1}, a')) $\;
        Perform gradient descent step on $(Y_{t}^{CDQL} - Q_{\theta_i}(s_t, a_t))^2$\;
        Update target networks parameters\;
        $\theta_{i}' \leftarrow \tau *\theta_{i} + (1-\tau) * \theta_{i}'$ \;
        Update $epsilon$ \textbf{if} $\epsilon > \epsilon_{min}$\;
        $ \epsilon *= \epsilon_{decay}$ 
        }
    }
     \label{cddqn}
    
     \caption{Clipped Double Q-Learning }
     
\end{algorithm}
\subsection{State space selection}\label{AA}
The state space is composed by two terms. The first one corresponds to the number of attached UEs' ratio in each BS and the second term is the Resource Block Utilization (RBU) vector. Thus, the state $S(t)$ will be represented by the concatenation of both metrics as: 

\begin{equation}
    S(t) = \begin{bmatrix} \mathbf{U(t)} & \mathbf{P(t)} \end{bmatrix}
\end{equation}
where $\mathbf{U}(t)$ corresponds to a $M_T$-length vector comprised by the ratio of UEs attached to each cell $i$. 
\begin{equation}
    \mathbf{U}(t) = \begin{bmatrix} u_1(t) & ... & u_{M_T}(t) \end{bmatrix}
\end{equation}
where $\Psi_{\Gamma_{i}}$ is the total of UEs attached to cell $i$  and $u_i = \dfrac{\Psi_{\Gamma_{i}}}{N_T}$

Additionally, $\mathbf{P}(t)$ corresponds to a $M_T$-length vector conformed by the RBU for each BS at time $t$ as:  
\begin{equation}
    \mathbf{P}(t) = \begin{bmatrix} p_1(t) & ... & p_M(t) \end{bmatrix}
\end{equation}

Given the fact that resource allocation in each cell is reported every TTI, which is a smaller time frame compared with the observable time interval of our agent (1s), we consider the resource block utilization in each TTI as a discrete random variable. Thus, we can  model the resource block utilization during the observation time $p_i$ as the expected value of the resource block utilization for each TTI $p_{tti}$. 
\begin{equation}
    p_i = \mathbb{E}[p_{tti}]
\end{equation}

\subsection{Reward}\label{AA}
The total reward function is calculated based on Quality of Service (QoS) parameters  and the load of each BS in terms resource block utilization as follows: 

\begin{equation}
    R_T(t) =  w_1*R_{D}(t) + w_2*R_{RB}(t) + w_3*R_{CQI}(t)
    \label{rew_total}
\end{equation}
Here $R_{D} \in [-1,1]$ is the reward related to delay constrains, $R_{RB} \in [-1,1]$ is the reward related to resource allocation usage and $R_{CQI} \in [-1,1]$ corresponds to a reward measuring the quality of the modulation used by the users in the network. Finally, $w_1,w_2,w_3$ are the weights for each individual reward. 

\begin{figure*}
\center
  \includegraphics[scale=0.50]{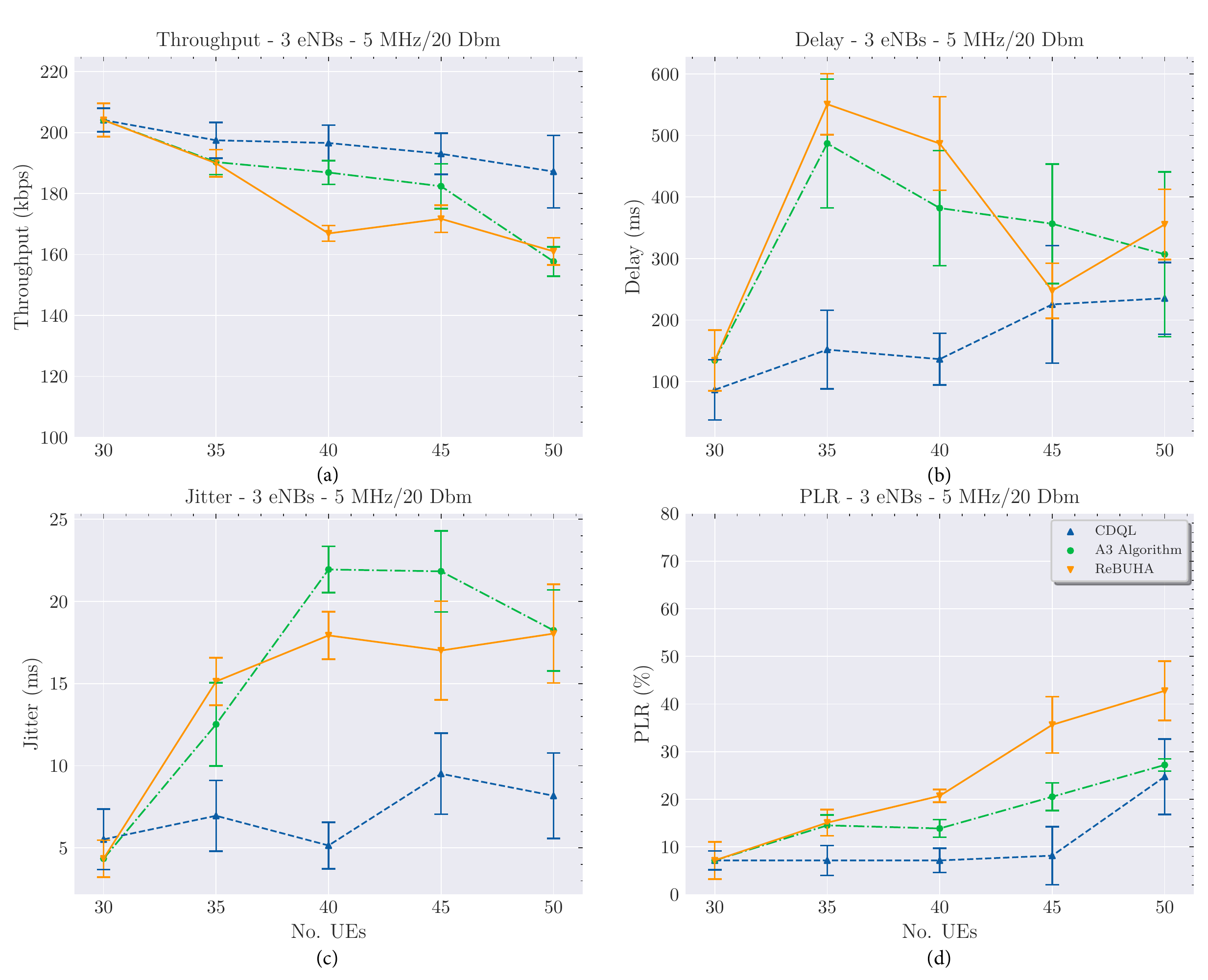}
  \caption{Performance metrics of CDQL, A3 and ReBuHA algorithms. (a) Throughput, (b) Delay, (c) Jitter and (d) PLR }
  \label{metrics}
\end{figure*}

The first component of the proposed reward can be defined as: 

\def\delequal{\mathrel{\ensurestackMath{\stackon[1pt]{=}{\scriptstyle\Delta}}}}

\begin{equation}
   R_{D}(t)\delequal \frac{1}{N_T} \sum_{i=1}^{N_T}\mathbb{1}\cdot\delta_{i}(t)  
\end{equation}
and, 
\begin{equation}
   \delta_{i}(t) =
    \begin{cases}
      -1 & \text{if $\exists \Psi_i \notin \Psi_C$ ,}\\
      \daleth(D_{avg}) & \textit{otherwise}
    \end{cases}       
\end{equation}
where $\Psi_i$ represents the $i$th UE and $\Psi_C$, $\Psi_C\subset \Psi$ is the set of connected UEs to any BS $\in \Gamma$. On the other hand, $\daleth(D_{avg})$ is a sigmoid function defined as follows: 

\begin{equation}
   \daleth(D_{avg}) = 1 + \dfrac{c}{1+ e^{-o(D_{avg} - \mathcal{F})}}
\end{equation}
Here $c$ establishes the upper bound of the slope, $o$ adjusts the slope of the sigmoid and $\mathcal{F}=2/3*PDB$ controls the target packet delay. $PDB$ is the Packet Delay Budget which according the type of traffic used in the network.

The objective of the first term is to reward each UE's average latency based on the PDB of a defined packet type. As can be seen, we penalize the cases if a UE gets disconnected from the network as a result of the load balancing decision. 

The second component of the proposed reward can be defined as:
\begin{equation}
R_{RB}(t)  \delequal 1 + \dfrac{c}{1+ e^{-a(max(P(t)) - \mathcal{D})}}
\label{second_rew}
\end{equation}
where $P(t)$ corresponds to a $M_T$-length vector composed of the resource block utilization of each BS at time t. The term $max(P(t))$ allows selecting the most loaded BS to either penalize or to reward based on a predefined   threshold defined as $\gamma_{RB}$. As the maximum value of $P(t)$ decreases, higher reward is obtained. $c$ establishes the upper bound of the slope, $a$ adjusts the slope of the sigmoid as in equation \ref{second_rew} and $\mathcal{D}$ controls the target data utilization.

Finally, the third component is defined by:
\begin{equation}
   R_{CQI}(t)\delequal \frac{1}{N_T} \sum_{i=1}^{N_T}\mathbb{1}\cdot\Omega_{i}(t)  
\end{equation}
where, 
\begin{equation}
   \Omega_{i}(t) =
    \begin{cases}
      -1 & \text{if $\Psi_{i_{CQI}}(t) < 6$,}\\
      0 & \text{if  $ 7 \leq \Psi_{i_{CQI}}(t) \leq 9$    ,}\\
      1 & \textit{otherwise}
    \end{cases}       
\end{equation}

$\Psi_{i_{CQI}}$ corresponds to the CQI measurement of the $i$th UE.
The objective of this term is to reward or penalize each UE's CQI based on the fact that higher CQI will be translated in a higher modulation scheme and thus a more efficient usage of the resource block allocation.
\begin{figure*}
\center
  \includegraphics[scale=0.55]{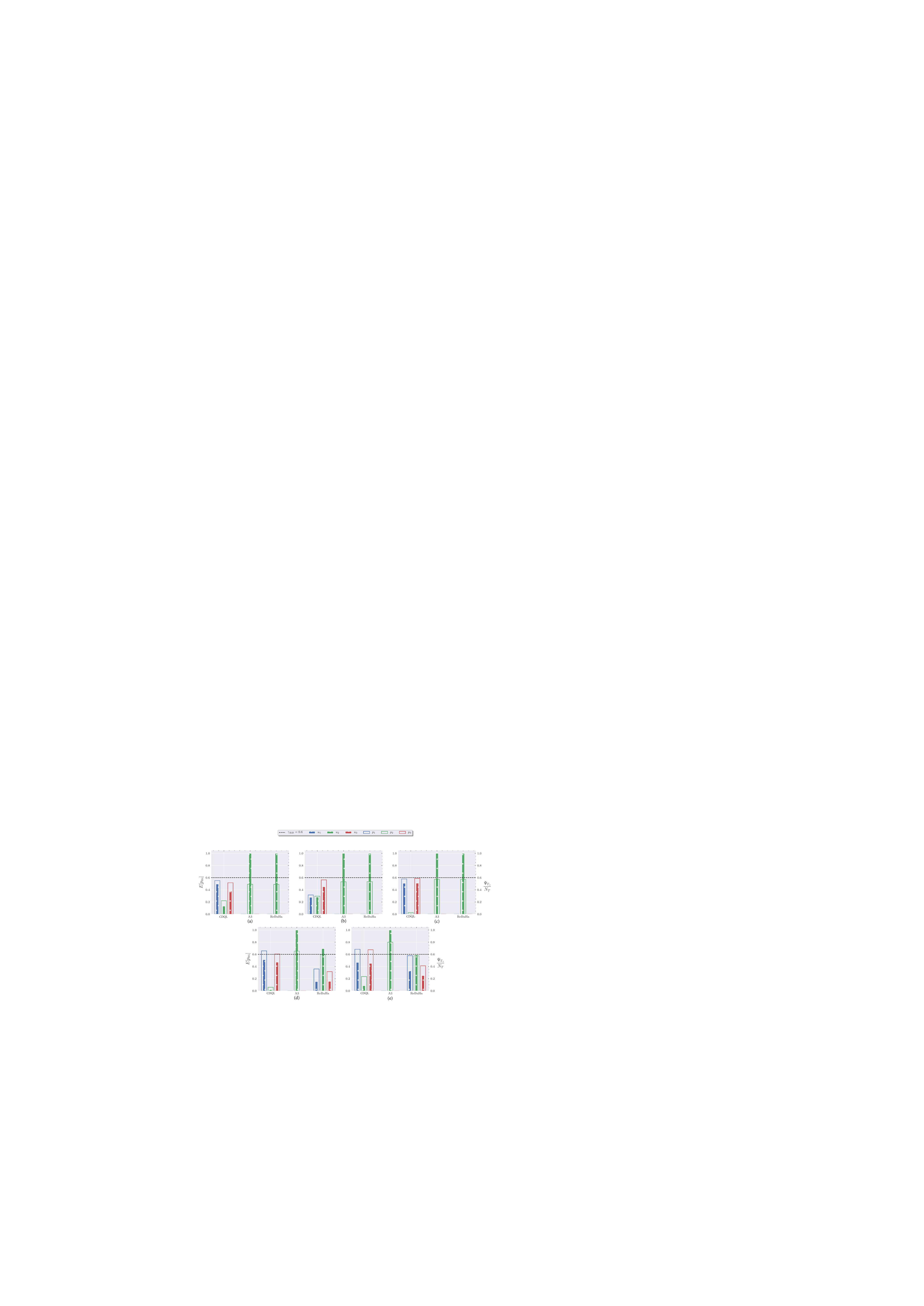}
  \caption{Average BS resource block utilization (RBU) and UE per BS ratio after convergence of each strategy. (a) 30 UEs, (b) 35 UEs, (c) 40 UEs, (d) 45 UEs, (e) 50 UEs}. 
  \label{rbu}
\end{figure*}

\subsection{Baseline: Resource Block Utilization based Handover Algorithm (ReBUHA)}\label{AA}
In this work, we consider a baseline algorithm named ReBUHA in addition to the classical A3 handover algorithm to perform a comparison of our results. ReBUHA algorithm replaces the A3 handover which mainly relies on received power (see eq. \ref{a3}) with a resource allocation awareness method to trigger the handover procedure.
\begin{algorithm}
 \KwData{RBU information of all BSs}
 \KwResult{Resource allocation based handover}
 
 \eIf{all $p_i(t) \in \mathbf{P}(t):  p_i(t) > \gamma_{RB}$}{
  break\;}
  {
   \ForEach{$p_{j \neq i}(t) \in \mathbf{P}(t)$ }{
        Check if UE $\Psi_i$ is attached to an overloaded BS\;
        \eIf{$p_{j}(t) < \gamma_{RB}$ }
            {trigger handover to cell $j \in \Gamma$\;}
            {break\;}
        }
   }
   
 \caption{ReBUHA algorithm}
 \label{baseline}

\end{algorithm}
As seen in Algorithm \ref{baseline} $\mathbf{P}(t)$ is $M_T$-size vector where each element corresponds to the ratio of RBs used in the $m^{th}$ BS at the time t. Every $t = 1s$, the algorithm will use $\mathbf{P}(t)$ information and look for handover opportunities in a centralized way. $\gamma_{RB}$ corresponds to a resource block utilization threshold defined as the ratio of usage in which the handover algorithm will look for opportunities for handover.  

\section{Performance Evaluation}
\subsection{Simulation Setting}\label{AA}

\begin{table} 
\caption{Network settings}
\begin{center}
\resizebox{\columnwidth}{!}{
\begin{tabular}{c c} 
\hline
\textbf{Parameter}&\textbf{Value} \\
\hline

Inter-site Distance & {720 m} \\
$M_T$ & { 3 } \\
$N_T$ & { 30,35,40,45,50 } \\
Center Frequency & {2 MHz} \\
System Bandwidth & { 5 MHz (25 resource blocks)} \\ 
Pathloss Model & { Log Distance Propagation Loss Model } \\
\textbf{} & { 95 + 27 $log_{10}(distance[km])$ } \\
BS antenna height & { 30 m } \\
UE antenna height & { 1.5 m} \\
Max Tx power & {20 dbm } \\  
MAC scheduler & { CQA scheduler} \\
User distribution & { Stationary and uniformly distributed} \\
\hline
Traffic Model & { Conversational video (live streaming) and Poisson } \\
\textbf{} & { Packet payload size = 250 Bytes } \\
\textbf{} & { Interval = 10 ms } \\
\textbf{} & { Packet delay budget = 150 ms } \\
Handover algorithm & { A3-event based } \\
 & { Time to trigger = 8 ms } \\
  & { Hysteresis =  2 dBm } \\
\hline
\end{tabular}
\label{net_settings}
}
\end{center}
\vspace{-4mm}
\end{table}
The simulations are performed by using the discrete network simulator ns-3 \cite{Baldo2011}. In Table \ref{net_settings} and Table \ref{q_settings} we provide the settings utilized in our simulations and the RL parameters, respectively. Each BS is positioned with an inter-site distance of 720 meters. Five different scenarios are tested under the proposed algorithms with 30, 35, 40, 45 and 50 UEs. Each scenario is designed by distributing a percentage of the total users on the edge of each cell in a random disc and the rest uniformly allocated throughout the coverage of the middle BS. 
We initialize the simulations by attaching all UEs to the BS that sits in between 2 BSs and then according to the policy of each strategy, handover is triggered or not. The traffic is a mixture of 20 UEs using CBR, and the rest following Poisson arrivals. For the case of Poisson traffic, we use a small payload of 32 bytes with a traffic load of 0.1 Mbps, meanwhile for the CBR which emulates video traffic we use a larger payload of 250 bytes with an interval of 10 ms. Simulation results are collected by averaging 15 simulations per scenario. Each simulation consists of 150 episodes with 50 iterations per episode. OpenAI Gym is used as the interface between ns-3 and our agent \cite{Gawowicz2018}. Lastly, we consider each parameter of our reward of equal importance, thus the weights of equation \ref{rew_total} are equal to 1.

\begin{table} 
	\caption{RL environment settings}
	\begin{center}
		\resizebox{\columnwidth}{!}{
		\begin{tabular}{c c} 
			\hline
			\textbf{Parameter}&\textbf{Value} \\
			\hline
			Number of iterations/episode & {50} \\
			Number of episodes &  {150  } \\
			Gym environment step time & { 1s } \\
			Batch size &  {32}  \\
			\hline
			CDQL & { $\o$ set  $\{-9,-6,-3,0,3,6,9\}$ dBm} \\
			\textbf{} & {$w_{1},w_{2},w_{3} = 1$}\\
			\textbf{} & {$\mathcal{F}=2/3*PDB$ where $PDB= 150$ ms, $c=-2 $, $o=75$}\\
			\textbf{} & {$\gamma_{RB} = 0.6$, $a = 20$}\\
			\textbf{} & {Optimizer :  Adam $(1e-3)$ }\\
			\textbf{} & {Number of hidden layers $(N_h) = $2 }\\
			\textbf{} & {Loss function : Huber Loss $(clip delta = 1.0)$ }\\
			\textbf{} & {Update target model type : Polyak averaging }\\
			\textbf{} & {$\gamma=0.95$, $\epsilon=1.0$, $\epsilon_{min} = 0.001$, $\epsilon_{decay} = 0.995$}\\
			\hline
		\end{tabular}
		\label{q_settings}
	}
\end{center}
\end{table}

\subsection{Simulation Results}\label{AA}
To assess the performance of our proposed scheme, we present throughput, delay, jitter, packet loss ratio (PLR) with 90\% confidence interval as well as the convergence of the machine learning algorithm.  

Figure \ref{cddqn_learning} shows the learning performance achieved by CDQL for the simulated scenarios. The trend shows how the reward value per episode converges in all cases and the converged reward value becomes lower as the number of UEs increase. This is due that some of the KPIs that are tracked by our agent are affected by the increment of the number of users.
Figure \ref{metrics}. presents throughput, end-to-end delay, jitter, and packet loss ratio (PLR). In Figure \ref{metrics}(a) it can be seen that our algorithm achieves an average improvement of 6.1\% and 9.5\% in throughput in comparison with the A3 and ReBuHa algorithms, respectively. Similarly, the other figures (b, c, d and e) show a noticeable improvement with respect to the baselines with a gain of 49.8\% and 52.9\% in terms of delay, 55\% and 51\% in terms of jitter and 34\% and 55.2\% in terms of PLR for the cases of the A3 and ReBuHa algorithms, respectively. Note that, the delay results of ReBuHa algorithm decreases at 45 UEs in comparison with A3. This is because, after 45 UEs the algorithm triggers its handover procedure as the middle BS surpasses the resource block utilization threshold predefined by the value of  $\gamma_{RB}$ . However, metrics such as PLR continue to worsen reassuring our thesis that in high traffic scenarios a closed track of QoS metrics is needed. 
Fig. \ref{rbu} presents the components of the vectors $\mathbf{P}$ and $\mathbf{U}$ at the end of the simulated scenarios. It can be seen how the A3 handover algorithm does not trigger in any of the scenarios by keeping all the UEs attached in the middle BS. The latter occurs because none of the UEs comply with the event trigger condition of the A3 algorithm. Furthermore, it is noticeable that in Figure \ref{rbu} (d) the ReBuHa algorithm starts distributing the load in agreement with the behavior described in the previous figure. For our proposed algorithm it is observable that it distributes the UEs over the BSs based on choosing the "best" CIO value per BS that will maximize our agent's reward objective function. 
Note that the proposed algorithm for (d) and (e) does not meet the target resource block utilization goal which is established by $\gamma_{RB}$. This behavior can be explained based on the needed minimization of not only the resource block utilization in the network but also improvement of QoS metrics as well.
\begin{figure}[h]
    \centering
  \includegraphics[scale=0.9]{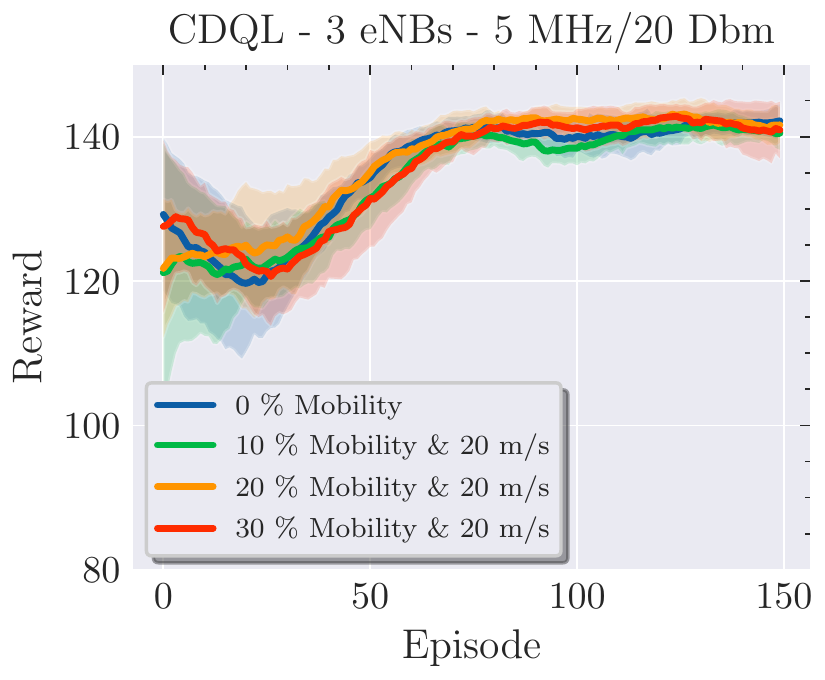}
  \caption{Learning performance of CDQL for 30 UEs and 0\%, 10\%, 20\% and 30\% of mobile users with UE's speed of 20 m/s. }
  \label{cddqn_mbile}
\end{figure}
\begin{figure*}
\center
  \includegraphics[scale=0.65]{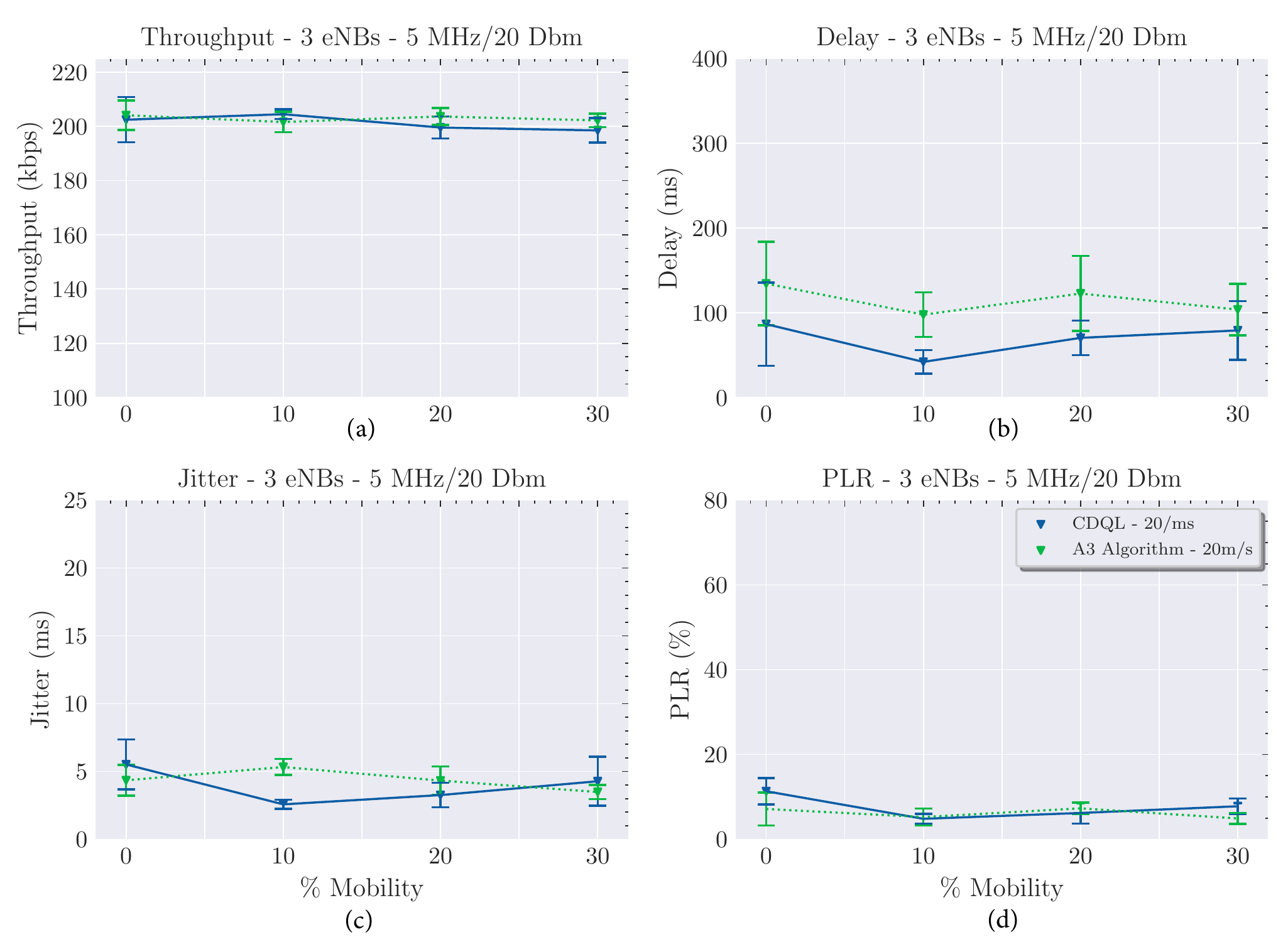}
  \caption{Different performance metrics of CDQL and A3 algorithms for different mobility percentages and UE's speed of 20 m/s. (a) Throughput, (b) Delay, (c) Jitter and (d) PLR }
  \label{metrics_mobile}
\end{figure*}

Additionally, we present the performance of our scheme under mobility. We consider 30 UEs where 10\%, 20\% and 30\% of the total UEs in the network are mobile and they use random walk with a speed of 20 m/s. Figure \ref{cddqn_mbile} shows that, similar to the non-mobility scenario (represented as 0\%), our scheme converges for mobile scenarios in a similar fashion. Furthermore, we show in Figure \ref{metrics_mobile} throughput, end-to-end delay, jitter, and PLR for non-mobile and mobile scenarios. Note that we only presented A3 Algorithm and CDQL since for such number of users the ReBuHa algorithm behave identical as the A3. In terms of throughput and PLR both CDQL and A3 algorithm perform similar. For delay, the proposed CDQL offers slightly lower latency than A3 algorithm. More specifically, our scheme is able to achieve an improvement of 64\%  respecting delay with no considerable difference in the other KPIs. 
Finally, it is worth to mention that our scheme shows its real potential when the number of users increases, in other words, when the scarcity of resources increases. We chose the A3 best performance scenario (30 UEs) to show the steady behavior of our scheme. 

\section{Conclusions }
In this paper, we presented a Clipped Double Q-Learning strategy that performs load balancing with awareness of QoS metrics. As main difference from previous works, our RL method uses a state-of-the-art algorithm and a QoS-aware load balancing approach enhancing the overall KPI metrics such as throughput, delay, jitter, and packet delivery ratio. We  compared our proposed scheme with two baselines: the traditional A3 handover algorithm and a resource block utilization based handover scheme, named ReBuHa. The results showed an average improvement up to 6.1\% and 9.5\% in terms of throughput, 49.8\% and 52.9\% in terms of delay, 55\% and 51\% in terms of jitter and 34\% and 55.2\% in terms of PLR, in comparison to the A3 and ReBuHa algorithms, respectively. Additionally, we evaluated the performance of our scheme under mobility. The results revealed the importance of performing load balancing while maintaining latency and CQI metrics.  

\section{Acknowledgment }
This research is supported by the 5G ENCQOR program and Ciena. 

\bibliography{biblio.bib}{}
\bibliographystyle{IEEEtran}

\end{document}